\documentclass[10pt, a4paper]{article}

\usepackage{amsmath}
\usepackage{amssymb}
\usepackage{amsfonts}
\usepackage{amsthm}
\usepackage{latexsym}
\usepackage{color}
\usepackage{epsfig}
\usepackage{natbib}
\usepackage{hyperref}

\newcommand{\proglang}[1]{{\tt #1}}       
\newcommand{\pkg}[1]{{\tt #1}}       
\newcommand{\code}[1]{{\tt #1}}       

\newcommand{\blanco}[1]{ }

\def\d{\, \mathrm d}                                
\def\Var{\mathop{\rm Var}\nolimits}

\newcommand{\ve}[1]{\boldsymbol{#1}}

\def\PP{{\cal P}}

\def\bea{\begin{eqnarray*}}
\def\eea{\end{eqnarray*}}
\def\be{\begin{equation}}
\def\ee{\end{equation}}
\def\bean{\begin{eqnarray}}
\def\eean{\end{eqnarray}}
\def\barr{\begin{array}}
\def\earr{\end{array}}
\def\bdes{\begin{description}}
\def\edes{\end{description}}
\def\bi{\begin{itemize}}
\def\ei{\end{itemize}}

\def\Bl{\Bigl}
\def\Br{\Bigr}

\def\Ex{\mathop{\rm I\!E}\nolimits}

\def\R{\mathbb{R}}

\def\argmax{\mathop{\rm arg\,max}}

\def\Var{\mathop{\rm Var}\nolimits}

\def\ed{\end{document}}

\begin{document}

\renewcommand{\baselinestretch}{1.15}

\title{\vspace*{-1cm}Selection models with monotone \\ weight functions in meta analysis}
\author{Kaspar Rufibach\thanks{Biostatistics Unit, Institute for Social and Preventive Medicine, University of Zurich, Hirschengraben 84, CH-8001 Zurich, Switzerland.
e-mail: kaspar.rufibach@ifspm.uzh.ch, Phone: +41-44-634-4643, Fax: +41-44-634-4386} \\
University of Zurich}

\date{\today}

\maketitle

\begin{abstract}
Publication bias, the fact that studies identified for inclusion in a meta analysis do not represent all
studies on the topic of interest, is commonly recognized as a threat to the validity of the results of a meta analysis.
One way to explicitly model publication bias is via selection models or weighted probability distributions.
We adopt the nonparametric approach initially introduced by \cite{dear_92} but impose that the weight function $w$
is monotonely non-increasing as a function of the $p$-value. Since in meta analysis one typically only has few studies or ``observations'', regularization
of the estimation problem seems sensible. In addition, virtually all parametric weight functions proposed so far in the
literature are in fact decreasing. We discuss how to estimate a decreasing weight function in the above model and illustrate the
new methodology on two well-known examples. The new approach
potentially offers more insight in the selection process than other methods and is more flexible
than parametric approaches. Some basic properties of the log-likelihood function and computation of a $p$-value quantifying the evidence against
the null hypothesis of a constant weight function are indicated. In addition, we provide an approximate selection bias adjusted profile likelihood
confidence interval for the treatment effect.
The corresponding software and the datasets used to illustrate it are provided as the \proglang{R} package \pkg{selectMeta} \citep{selectMeta}.
This enables full reproducibility of the results in this paper.
\end{abstract}


\newtheorem{theorem}{Theorem}[section]
\newtheorem{lemma}[theorem]{Lemma}

\newenvironment{Theorem}{\begin{theorem}\sl}{\end{theorem}}
\newenvironment{Lemma}{\begin{lemma}\sl}{\end{lemma}}





\section{Introduction} \label{intro}
Meta analysis has become a widely used technique for synthesizing evidence from different studies,
see e.g. \cite{sutton_08} for an overview over recent developments.
Publication bias, i.e. the fact that studies identified for inclusion in a meta analysis, do not represent all
studies on the topic of interest, is commonly recognized as a threat to the validity of the results of a meta analysis.
Overviews how to prevent, assess, and adjust for publication bias are provided in \cite{sutton_00}, \cite{macaskill_01}, or \cite{rothstein_05}.

Numerous tools to detect publication bias in meta analysis have been developed, see \citet[Chapters 5-11]{rothstein_05}
for an excellent overview of the current state-of-the-art. 

If one seeks to assess selection bias one
typically requires some model for the sampling behavior of the observed effect sizes that explicitly incorporates
the selection process \citep{hedges_05}. It is hence useful to distinguish two parts of such a model: the effect size
part and the selection part. The former specifies what the distribution of the effect sizes would be if there were no
selection whereas the latter explicitly models how the effect size distribution is modified by the selection process.
Two different classes of explicit selection models have been proposed so far for meta analysis \citep{hedges_05}.
The first class depends on the effect size estimate, such as relative risk or odds ratio, and the corresponding standard
error separately, see \cite{copas_99}, \cite{copas_00}, \cite{copas_01}, and the implementation in the \proglang{R} package
\pkg{copas} \citep{carpenter_09}. This type of model is typically denoted ``Copas selection model''.
In the second class the weight is assumed to depend on the effect size only via the $p$-value associated with the study, see \cite{hedges_84},
\cite{iyengar_88}, \cite{hedges_92}, \cite{dear_92}, \citet[p. 149]{hedges_05} or \cite{copas_08} for a test on selection bias that is robust against any
form of selection function. The rationale to make the weight function depending on
$p$-values, or equivalently on the standardized effect size, exclusively is that often, decisions about conclusiveness of medical research results
are based on statistical significance (only).

More specifically, following the development in \cite{hedges_05}, let $Y^*$ be a random variable with density $f(y|\theta, \sigma)$
representing the effect estimate before selection, typically assumed to follow a normal distribution.
Denoting the weight function by $w(y)$, the weighted density of the observed effect estimate $Y$ is then given by
\bea
    g(y |\theta, \sigma) &=& \frac{f(y|\theta, \sigma) w(y)}{\int f(y|\theta, \sigma) w(y) \d y}.
\eea Whenever the weight function $w$ is not constant, the sampling distribution of the observed effect size $Y$ differs from that of the
unselected effect size $Y^*$ and this difference, i.e. the shape of $w$, is a way of describing selection bias.


Now, if larger values of $Y^*$ are more likely to be observed than smaller values, $w(y)$ is a monotone non-decreasing function of the effect size $y$.
Considering $w$ on the scale of $p$-values this implies that $w(p)$ as a function of the $p$-value is non-increasing, meaning that smaller $p$-values are
more likely to be observed than larger $p$-values.
In this paper we propose a non-increasing estimate $\hat w(p)$ in the nonparametric normal model introduced by \cite{dear_92}.
Besides being a plausible assumption as elaborated above, nonparametrically estimating the weight function $w(p)$ and imposing a monotonicity constraint
has further advantages:
\bi
\item All parametric weight functions proposed in the literature are in fact non-increasing, see Section~\ref{monotonicity} for a brief discussion.
\item Typically, the number of studies that enter a meta analysis is small to moderate. For this reason, additional
regularization, such as monotonicity, and therewith constraining the parameter space, may lead to more
realistic but still flexible estimates of the weight function compared to the purely nonparametric approach by \cite{hedges_92}
and \cite{dear_92}, but without forcing a purely parametric model. This makes our approach less prone to misspecification.
See also the comment in \citet[p. 118]{hedges_88}.
\item Restricting the parameter space, or shape of the function as in our case, typically yields estimates with better performance, e.g. measured in terms of mean squared error, if in fact the
function to be estimated has the assumed shape, see e.g. \citet[p. 937]{kelly_89}.
\item In contrary to e.g. kernel estimators or the penalized monotone estimator of \cite{sun_97}
the estimator $(\ve{\hat w}, \hat \theta, \hat \sigma^2)$ defined below does not necessitate the choice of a smoothing or
penalty tradeoff parameter (or a prior) and is therefore fully automatic.
\item Weight functions are primarily proposed as an exploratory and informal means to assess the degree of
publication bias which may be present, see \citet[p. 240]{dear_92} or \citet[p. 431]{sutton_00}. Specifically, if there is no selection effect
at work, the former authors claim that the graphs of their estimated unconstrained weight function ``provide visual confirmation
of the lack of bias, demonstrating a seemingly random configuration of estimated weights.''
However, it is not without difficulty to identify the model without biased selection from the estimated weight
functions in sub-figures (a), (b), (c) of \citet[Figure 2]{dear_92}. As reveal our examples in Section~\ref{examples},
the monotonicity assumption typically yields more insight in the actual selection process.
\ei

In Section~\ref{setup} we derive the log-likelihood function in our nonparametric model and provide some properties of it whereas in
Section~\ref{monotonicity} we discuss different approaches to setup selection models and choose sensible selection functions.
Section~\ref{computation} elaborates on the computation of our proposed estimate. A discussion of statistical inference for the effect
$\theta$ and the random effects variance component $\sigma^2$ are provided in Section~\ref{inference}. Specifically, in this section
we sketch derivation of a profile likelihood confidence interval for the selection bias adjusted treatment effect $\theta$.
A way to quantify evidence against the null hypothesis of no selection is described in Section~\ref{quantifying evidence}.
The paper is concluded with the analysis of two well-known examples
and a discussion of the software package \pkg{selectMeta} \citep{selectMeta} that enables full reproducibility of the results presented in this paper.

\section{The log-likelihood function and its properties} \label{setup}
To fix ideas, assume that there are $n$ independent studies with normally distributed observed treatment effects
$Y_i$, $i = 1, \ldots, n$ where $\Ex(Y_i) = \theta$ and $\Var(Y_i) = \eta_i^2 = u_i^2 + \sigma^2$.
Here, $u_i^2$ is the known sampling variance in the $i$-th study (largely determined by the sample size in the $i$-th
study and therefore considered known) and $\sigma^2$ is a random effects component of variance representing
the heterogeneity in the population. Typically, it is assumed that the effects follow a normal distribution,
i.e. $Y_i \sim N(\theta, \eta_i^2)$ with realizations $y_i$. The two-sided $p$-values for the null
hypothesis $H_0: \theta = 0$ can then be computed in each study as
$p_i = 2\Phi(-|y_i|/u_i)$ and are, in accordance with the notation of \cite{dear_92},
considered to be ordered and denoted by $p_n, \ldots, p_1$, where $p_n$ is the smallest and $p_1$ the largest.
Furthermore, let $p_{n+1} = 0$ and $p_0 = 1$. Assume that the selection process is governed by the non-negative
weight function $w$ that assigns to an effect estimate the likelihood that it is observed. Then, the 
likelihood function of the observed effect sizes $\ve{y} = (y_1, \ldots, y_n)$, given the weight function $w$, the quantities $\theta$, $\sigma^2$, and $\ve{u} = (u_1, \ldots, u_n)$, amounts to
\bean
    L(\ve{y}| w, \theta, \sigma^2, \ve{u}) &=& \prod_{i=1}^n P(y_i | \text{$i$-th study is published}) \nonumber\\
    &=& \prod_{i=1}^n \frac{\phi\Bl((y_i-\theta)/\eta_i \Br)w(y_i)}{A_i(w, \theta, \sigma^2, \ve{u})} \label{eq: L}
\eean where we introduced the normalizing constant 
\bean
    A_i(w, \theta, \sigma^2, \ve{u}) &=& \int_{-\infty}^\infty \phi\Bl((y-\theta)/\eta_i \Br) w(y) \d y \label{def: Ai}
\eean and $\Phi$ as well as $\phi$, the cumulative distribution and density function of a standard normal distribution.
Now observe that in \eqref{eq: L} the unknowns are the function $w$ and the parameters $\theta$ and $\sigma^2$.
Numerous suggestions have been made to estimate these unknowns, where these proposals differ by the assumptions
they impose on the selection function $w$, see the discussion in Section~\ref{monotonicity}.

In this paper, as in \cite{dear_92} and \cite{hedges_92}, we posit that the weight function $w$
is a left-continuous step function of the $p$-value. In \cite{hedges_92}, the discontinuities of $w$ are
fixed at, say, ``psychologically motivated'' values, whereas \cite{dear_92} group the $p$-values in pairs and
assume equal values of $w$ for two adjacent observed $p$-values. Here, we adopt the latter approach noting that
the former model fits in our framework equally well. More specifically, the weight function
is, for $p \in [0, 1]$, defined as
\bea
    w(p) &=&
    \begin{cases}
    w_j & \text{ if } p_{2j-2} \ge p > p_{2j} \\
    w_k & \begin{cases}
          p_{n-1} \ge p > 0 & \text{ if } n \text{ odd} \\
          p_n     \ge p > 0 & \text{ if } n \text{ even}
          \end{cases}
    \end{cases} \label{def: w}
\eea where $j \ = \ 1 + \lfloor i/2 \rfloor = 1, \ldots, k$ and $k$ is the number of categories that are built from
the initial $p$-values through pairing.
For reasons of identifiability one is not able to set up a likelihood assuming a piecewise
constant weight function without some sort of grouping of the $p$-values (see \citealp[Section 2.3.7]{sutton_00} and \citealp[Section 2]{dear_92}).
For the description of a ``pure selection model'' and the necessary modifications of the problem, we refer to \cite{sun_97}.

The weight function on the scale of the outcomes $y$ writes as:
\bean
    w(y) 
    &=& w_j1\{-u_i \Phi^{-1}(p_{2j}/2) > |y| \ge -u_i \Phi^{-1}(p_{2j-2} / 2)\}  \ \ \ \text{ for } i = 1, \ldots, n, \ j = 1, \ldots, k.  \label{wi}
\eean To see this, note that if the $p$-value in study $i$ gets the weight $w_j$ assigned, this $p$-value
is computed for a test statistic $|y_i|/u_i$ and equal to $p_h = 2\Phi(-|y_i|/u_i)$,
what gives $|y_i| = -u_i \Phi^{-1}(p_h/2)$.
Plugging in this form for the weight function $w$ into \eqref{eq: L} and taking the log yields the
final weighted log-likelihood function $l(\ve{w}, \theta, \sigma^2)$ for the parameter vector
$(\ve{w}, \theta, \sigma^2) \in \R^{k + 2}$. This log-likelihood was initially derived in \cite{dear_92}.
However, here and in the appendix we summarize its detailed development and discuss some properties and additional
computational facts.
For the log-likelihood function we get
\bea
    l(\ve{w}, \theta, \sigma^2)
    &=& -(n/2)\log(2\pi) + \sum_{j=1}^k \lambda_j \log w_j - \sum_{i=1}^n \log\eta_i  -\frac{1}{2}\sum_{i=1}^n \Bl( \frac{y_i-\theta}{\eta_i}\Br)^2 -  \sum_{i=1}^n \log A_i
\eea where $A_i, i = 1, \ldots, n$ are the normalizing constants defined in \eqref{def: Ai}.
Straightforward computations for any $c > 0$ 
yield that the log-likelihood function can be written as
\bean
    l(c \ve{w}, \theta, \sigma^2) &=& l(\ve{w}, \theta, \sigma^2) + \log(c) \Bl\{\Bl(\sum_{j=1}^k \lambda_j\Br) - n\Br\} \nonumber \\
&=& l(\ve{w}, \theta, \sigma^2) + \log(c) (\lambda_1 - 1). \label{eq: ll c}
\eean
Two important observations can be made for $l$.
First, the quantities $\lambda_j$ should, in principle, correspond to the number of $p$-values in any interval
$(p_{2j}, p_{2j-2}]$, $j = 1, \ldots, k$, i.e. $\lambda_1 = 1, \lambda_j = 2$ for $j=2, \ldots, k-1$
and $\lambda_k = 1 + 1\{\mbox{if }n\mbox{ is odd}\}$. However, the choice $\lambda_1 = 1$ would imply by
\eqref{eq: ll c} that the maximizer $\ve{\hat w}$ of $l$ was not identifiable.
To overcome this problem, \citet[p. 239]{dear_92} advise setting $\lambda_1 = 2$, so that \eqref{eq: ll c}
simplifies to $l(c \ve{\hat w}, \theta, \sigma^2) \ = \ l(\ve{\hat w}, \theta, \sigma^2) + \log c$, making
$\ve{\hat w}$ (1) identifiable but (2) estimated with a slight negative bias. For reasons of simplicity we choose $\lambda_1 = 2$
in the examples in Section~\ref{examples}. In the code collected in \pkg{selectMeta} \citep{selectMeta} the weight $\lambda_1$ can be set to an arbitrary value.

Second, \eqref{eq: ll c} entails that we must have $\ve{\hat w} \in (0, 1]$ since $\hat w_1 = 1$ once $\lambda_1 > 1$ is chosen.
To see this, assume $\ve{\hat w}$ with $\hat w_1 < 1$ the largest element and choose $c = 1 / \hat w_1 > 1$. This yields
\bea
    l(c \ve{\hat w}, \theta, \sigma^2) &=& l((1, \hat w_2 / \hat w_1, \ldots, \hat w_k/\hat w_1)', \theta, \sigma^2) \\
&=&l(\ve{\hat w}, \theta, \sigma^2) - \log(\hat w_1)(\lambda_1 - 1) \ \ \text{ via \eqref{eq: ll c}}
\eea and thus
\bea
    l((1, \hat w_2 / \hat w_1, \ldots, \hat w_k/\hat w_1)', \theta, \sigma^2) &>& l(\ve{\hat w}, \theta, \sigma^2)
\eea if $\lambda_1 > 1$.
However, as discussed in \cite{dear_92}, the actual selection probability is typically less than 1 for all
studies under consideration since some selection is going on for all studies, or rather the corresponding $p$-value.
As a consequence, the estimated weights are only relative. Since no information is available on the $p$-values
of the unpublished studies, one is not able to estimate the weight function directly.

The primary goal of this work is to specialize the approach of \cite{dear_92} to a monotone selection function $w$. Thus, we followed the framework developed in the latter paper, to enable straightforward comparison of the newly introduced monotone weight function
to the existing approaches and thus made the weight function $w(p)$ depending on two-sided $p$-values. However, the entire framework can straightforwardly be adapted to
one-sided $p$-values.

Ideally, in order to apply standard algorithms to maximize a log-likelihood function one appreciated if it would be nicely behaved, i.e. strictly concave
and coercive. Unfortunately, this is in general not the case for $l(\ve{w}, \theta, \sigma^2)$. Instead, plots of $l$ as a function of one of its arguments reveal that
it is not necessarily concave in $w_j, j = 1, \ldots, k$ and $\sigma$. However, these same plots strongly indicate that $l$ is at least unimodal with a
unique maximum, although we are not able to provide a formal proof of this property or some (even stronger) surrogate, like e.g. log-concavity of $l$.
Assuming that in fact $l$ were unimodal, Lemma~\ref{lem: l} below would then imply that a maximizer always exists.
In addition, the expression of the likelihood in the lemma also sheds some light on the peculiar structure of $l$.
To state Lemma~\ref{lem: l}, let $\ve{\rho} = (\ve{w}, \theta, \sigma^2)$.
\begin{lemma}\label{lem: l}
Assume that $n \ge 3$, $w_k > 0$, and $\lambda_j < n$ for all $j$.
Then, the log-likelihood function $l$ is continuous as a function of $(\ve{w}, \theta, \sigma^2)$ and coercive when one or more coordinates approach
the boundary of the domain, i.e. if $\Vert\ve{\rho}\Vert \to \infty$ and/or if $w_j \to 0$ for at least one $j < k$, then $l \to -\infty$.
\end{lemma}

Note that $l$ remains finite if $\sigma \to 0$ and all other arguments are kept fixed.

\paragraph{Proof of Lemma~\ref{lem: l}.} First, note that for a fixed $i$,
\bea
     \log A_i &=& \log(w_1 H_{i1} + \ldots + w_k H_{ik}) \\
              &=& \log(w_1) + \log(H_{i1} + \ldots + w_k H_{ik} / w_1) \\
              &=& \Bl(\sum_{j=1}^k \log w_j\Br) + \log\Bl\{\sum_{j=1}^k H_{ij} \Bl(\Pi_{l=1, l\ne j}^k w_l\Br)^{-1}\Br\}.
\eea Using this, the log-likelihood function can be written as
\bean
    l(\ve{w}, \theta, \sigma^2) &=& -(n/2)\log(2\pi) + \sum_{j=1}^k \lambda_j \log w_j - \sum_{i=1}^n \log\eta_i  -\frac{1}{2}\sum_{i=1}^n \Bl( \frac{y_i-\theta}{\eta_i}\Br)^2 -  \sum_{i=1}^n \log A_i \nonumber \\
    &=& -(n/2)\log(2\pi) + \sum_{j=1}^k (\lambda_j-n) \log w_j- \frac{1}{2}\sum_{i=1}^n \log(u_i^2+\sigma^2) \nonumber \\
    && \hspace*{0.75cm} - \frac{1}{2}\sum_{i=1}^n (y_i-\theta)^2(u_i^2 + \sigma^2)^{-1} - \sum_{i=1}^n\log\Bl\{\sum_{j=1}^k H_{ij} \Bl(\Pi_{l=1, l\ne j}^k w_l\Br)^{-1}\Br\}. \label{eq: loglik simple}
\eean
Let $\ve{\rho}_r$ be a sequence of vectors such that $\Vert \ve{\rho}_r \Vert \to \infty$ as $r \to \infty$.
From the definition of $H_{ij}$ in Appendix~\ref{wloglik} it is clear that $H_{ij} \in [0, 1]$.
The assumption $w_k > 0$ entails that at least one $H_{ij}$ is different from 0. From \eqref{eq: loglik simple} it is then not difficult to see that
\bea
    l(\ve{w}_r, \theta_r, (\sigma^2)_r) &\to& - \infty \ \mbox{ as } \ r \to \infty
\eea for either combination of possibilities, i.e. $w_{r, j} \to \infty$ for one or more $j's$
and/or $|\theta_r| \to \infty$
and/or $(\sigma^2)_r \to \infty$. Representation \eqref{eq: loglik simple} also illustrates the continuity of $l$.
Now, from \eqref{eq: loglik simple} we can derive that
\bea
    l(\ve{w}, \theta, \sigma^2) &=& -(n/2)\log(2\pi) + \lambda_1 \log w_1 + \sum_{j=2}^k (\lambda_j-n) \log w_j- \frac{1}{2}\sum_{i=1}^n \log(u_i^2+\sigma^2) \nonumber \\
    && \hspace*{0.75cm} - \frac{1}{2}\sum_{i=1}^n (y_i-\theta)^2(u_i^2 + \sigma^2)^{-1} - \sum_{i=1}^n\log\Bl\{H_{i1} w_1 \Bl(\Pi_{l=2}^k w_l\Br)^{-1} + \sum_{j=2}^k H_{ij} \Bl(\Pi_{l=2, l\ne j}^k w_l\Br)^{-1}\Br\}.
\eea
Without loss of generality assume that $\Vert \ve{\rho}_r \Vert \to \infty$ or to some constant, but that $w_{r, 1} \to 0$. The above representation then
readily implies that $l(\ve{w}_r, \theta_r, (\sigma^2)_r) \to - \infty$. \hfill $\Box$


\section{Monotone selection function} \label{monotonicity}

For a thorough review of weight functions $w$ proposed in the literature for meta analysis we refer to
\citet[Section 2]{sutton_00}. The spectrum ranges from (1) fully parametric proposals as in \cite{iyengar_88}, the
weight function proposed in the comment to \cite{iyengar_88} by Hedges, or those in \citet[Section 3.2]{preston_04}
to (2) nonparametric models as those discussed in \cite{hedges_92} and \cite{dear_92}.
Many of these functions have also been considered in a Bayesian framework, see the discussion in
\citet[Section 2]{sutton_00} or \cite{silliman_97_jasa, silliman_97}.

In general, there is little empirical evidence to guide the choice of weight functions (\citealp[p. 119]{hedges_88}).
However, the literature generally agrees that weight functions that depend on the effect
size through the corresponding $p$-value should be non-increasing as a function of the $p$-value,
see \citet[p. 238]{dear_92}, \citet[p. 38]{iyengar_94}, or \cite{lee_01postcon}.
In the review by \citet[Section 2.3]{sutton_00} all eight weight functions that are discussed are in fact
monotone non-increasing.
On the other hand, \citet[p. 249]{hedges_92} argues that ``It is probably unreasonable to assume that much is known
about the functional form of the weight function.'' Combining these two demands we therefore propose to
adopt the approach by \cite{dear_92}, i.e. to use the log-likelihood function $l(\ve{w}, \theta, \sigma^2)$
developed in Section~\ref{setup}, but maximize it over the set
\bea
    \PP &=& \{(\ve{w}, \theta, \sigma^2) \ : \ 1 \ = \ w_k \ge \ldots \ge w_1, \theta \in \R, \sigma \ge 0\}.
\eea 
So, we aim at computing
\bean
    (\ve{\hat w}, \hat \theta, \hat \sigma^2) &=& \argmax_{(\ve{w}, \theta, \sigma^2) \in \PP} l(\ve{w}, \theta, \sigma^2). \label{eq: problem}
\eean For completeness, we also state the unconstrained problem of \cite{dear_92}
\bean
    (\ve{\hat w}_*, \hat \theta_*, \hat \sigma_*^2) &=& \argmax_{(\ve{w}, \theta, \sigma^2) \in [0, 1]^k\times\R\times[0,\infty)} l(\ve{w}, \theta, \sigma^2). \label{eq: problem unconstr}
\eean

To conclude this section we would like to point out \citet[p. 228]{givens_97}, for two reasons:
First, to the best of our knowledge these are the only authors who explicitly estimate a monotone non-increasing
weight function.
However, in a Bayesian context via rejection sampling. Second, they remark that
``Such a [monotonicity] constraint is much harder to put in place in the frequentist setting...''. Here, we close this gap
for the likelihood setup described above and provide corresponding \proglang{R} software, see Section~\ref{software}.

\paragraph{Sensitivity with regard to assumptions.}
As discussed in the introduction, explicitly modeling the selection function depending on the study $p$-value is only one way of adjusting for selection bias,
the most prominent alternative being the Copas selection model that makes selection depending on the effect size estimate $T$ and the corresponding
standard error $\sigma$ separately (\citealp{copas_99}, \citealp{copas_00}, \citealp{copas_01}, \citealp{hedges_05}, \citealp{carpenter_09}).
However, making the selection function depending on the $p$-value only has the longest history in meta analysis \citep[p. 149]{hedges_05}.
A potential constraint of the latter models is that they treat equally significant results in the same way, irrespective of the size of the underlying study
and the direction of the effect. Thus, a potential next step to generalize the approach proposed here is to setup a two-dimensional selection function
that is
\bi
\item non-increasing as a function of $p$-values and
\item non-decreasing as a function of the underlying study size.
\ei
A potential source of misspecification is an inappropriate choice of $w$'s shape. However, as elaborated in Section~\ref{intro},
monotonicity seems a very plausible assumption for a selection function and all proposed parametric approaches are in fact non-increasing.
Finally, in order to correct for publication bias, selection models must substitute assumptions for data that are missing.
In our scenario, we stipulate that the form of the unselected distribution of the effect size estimates is normal. However, \cite{hedges_96}
performed a large simulation study assessing robustness of estimation from selection models to misspecification of effect distribution and concluded
that, surprisingly, the procedure is rather robust in this regard \citep{hedges_05}.

Note that estimation in the Copas selection model is not free of difficulties and estimation can be impossible for certain parameter values
\citep{hedges_05}.

\section{Computational aspects} \label{computation}
Having formulated the problem \eqref{eq: problem} it remains to numerically compute $(\ve{\hat w}, \hat \theta, \hat \sigma^2)$.
As a consequence of the considerations in Section~\ref{setup}, properly maximizing $l$ is
somewhat delicate, even when looking at the unconstrained problem \eqref{eq: problem unconstr}.
Note that neither \cite{dear_92} nor \cite{hedges_92} discussed this aspect.
They both mention (\citealp[p. 240]{dear_92} and \citealp[p. 251]{hedges_92}) that a multivariate Newton-Raphson procedure
can be used to find the unconstrained maximum $(\hat {w_1}_*, \ldots, \hat {w_k}_*, \hat \theta_*, \hat \sigma_*^2)$.
To avoid inversion of the corresponding Hessian matrix \cite{dear_92} use an EM-type algorithm which is also discussed in
\cite{hedges_92}. Namely, one iterates optimization for one entry of $(\ve{w}, \theta, \sigma^2)$ at a time
using Newton-Raphson until convergence. We implemented both approaches and surprisingly, the one-entry-at-a-time
version turned out to be more stable and quick enough and has therefore been implemented in \pkg{selectMeta}, see Section~\ref{software} for details.

However, we did not find a way to generalize this approach to find our new, constrained estimator
$(\ve{\hat w}, \hat \theta, \hat \sigma^2)$ defined in \eqref{eq: problem}. In fact, to find this estimator we have to
maximize a (most likely) unimodal, coercive but generally non-concave function under constraints, a non-trivial global optimization
problem. The solution we present makes use of the so-called evolutionary global optimization via the differential evolution (DE)
algorithm, initially described in \cite{storn_97}. This algorithm is particularly well-suited to find the global optimum of a real-valued function
of real-valued parameters, such as our log-likelihood function $l$. Neither continuity nor differentiability is a necessary property of the target function
that is maximized by a DE algorithm (as a matter of fact, our $l$ above is differentiable).
An implementation of a DE algorithm is made available in the \proglang{R} \citep{R} package
\pkg{DEoptim} \citep{DEoptim}. The function \code{DEoptim} allows for unconstrained global maximization. To account for constraints,
as in our case given by the monotonicity assumption in the parameter set $\PP$, one simply integrates the constraint within the function to optimize by penalizing
deviations from the constraints with $-\infty$. Set up this way, the function \code{DEoptim} quickly and reliably delivers the
maximum of $l$ over $\PP$. For a description of the implemented software we refer to Section~\ref{software}.

\section{Statistical inference on $\theta$ and $\sigma^2$} \label{inference}
We agree with \citet[p.120]{hedges_88} when he says that ``Although I am enthusiastic
about the development of varied and realistic models for estimation under selection, I do not believe that estimates
from any one of these models should be taken too seriously.''
This approach of considering selection models a way of exploring the selection mechanism, but not to estimate
the parameters $\theta$ and $\sigma^2$, is further supported by \citet[p. 240]{dear_92} claiming that ``The procedure
presented here [meant is their selection model] is intended primarily as a means of informally exploring the degree
of publication bias which may have operated in the selection of studies contributing to a meta analysis. Inference
about $\theta$ and $\sigma^2$ should be considered secondary at this stage.''
or by \citet[p. 431/439]{sutton_00}: ``Clearly, it is far more desirable to alleviate the problem of publication bias
rather than try to model it analytically. [...] Hence, the weight function obtained is used to provide a visual
display of the relative weight function for the purposes of identifying publication bias,
and is not used to adjust the pooled estimate.''
If selection bias is suspected from looking at $\ve{\hat w}$ (or $\ve{\hat w}_*$) one should primarily focus attention on the
possible causes of bias, e.g. initiate a search for ``missing'' studies, rather than using the model
to adjust $\hat \theta$ and $\hat \sigma^2$ for publication bias.

However, to get a complete picture we would like to sketch a way of making selection bias adjusted inference for $\theta$.
As elaborated in the seminal paper by \cite{murphy_00} on profile likelihood in presence of an infinite-dimensional nuisance parameter,
ordinary profile likelihood inference may still be applicable if the entropy of the function class of the nuisance parameter is not too large.
When seeking inference for $\theta$, the nuisance parameters are $\sigma$ and the estimated monotone weight function, $w$. That the class
of monotone functions is not ``too large'' in terms of entropy and thus accessible for the approach by \cite{murphy_00}
is discussed in \cite{fan_00}. \cite{ghosh_07} uses this approach to provide inference on a one-dimensional parameter with an estimated monotone
nuisance function in the evaluation of a biomarker. Here, by appealing to the above profile likelihood arguments of \cite{murphy_00}, we get that
the likelihood ratio-based statistic for the parametric component $\theta$ will have a $\chi^2$ limiting distribution with one degree of freedom.
Based on this result we derive a selection bias adjusted profile likelihood confidence interval for $\theta$ which is implemented
as the function \code{DearBeggMonotoneCItheta} in \pkg{selectMeta}.
Note that this procedure is approximate and rigorous theoretical justification of this approach will be provided elsewhere.

\section{Quantifying the evidence against a constant weight function} \label{quantifying evidence}
Obviously, one would like to have a mean to quantify the evidence against the null hypothesis of a constant weight function $w$.
The only reference we are aware of that deals with a similar problem is \cite{woodroofe_99}. A monotone $w$ is considered, but the density of the effect sizes
is assumed to be entirely known what precludes application to our situation.

However, as an alternative we suggest a simulation procedure to get a $p$-value that quantifies the evidence against a constant weight function $w$,
based on our new monotone estimator. Before describing computation of such a $p$-value let us
introduce the density function $g$ of the distribution of $p$-values for a meta analysis with true
effect $\theta$, variance $u^2$, and random effect component $\sigma^2$:
\bean
f(p; \theta, \sigma, \eta) &:=& \frac{\sigma}{2 \eta} \frac{\phi\Bl[\{-\sigma \Phi^{-1}(p / 2) - \theta\} / \eta\Br] + \phi\Bl[\{\sigma \Phi^{-1}(p / 2) - \theta\} / \eta\Br]}{\phi\{\Phi^{-1}(p / 2)\}}, \label{pval dens}
\eean
where $\eta^2 = u^2 + \sigma^2$. This is the density generated by a test of the hypothesis $H_0: Y \sim N(0, \sigma^2)$
vs. $H_1: Y \sim N(\theta, \eta^2)$. Note that $f$ simplifies to denoted by $g(p)$ in \citet[p. 240]{dear_92}
for their choice $u=1, \sigma= 0$ of parameters.

The log-likelihood function $l$ does not depend on the $p$-values only, but also on the sign of the initial effect
size $y$. So when simulating $p$-values from $f$, to be able to compute the log-likelihood function, it is therefore not sufficient to generate
a sample of $p$-values from the density $f$ only (via numerical inversion of the
quantile function corresponding to $f$) but one also needs to randomly generate the signs of the corresponding ``observations'' $y$.
For each generated $p$-value $p$ and a Bernoulli random variable $Z \sim Ber(1/2)$ we therefore compute
$y^* = -u \Phi^{-1}(p/2)$ and set
\bea
    y &=& \begin{cases} y^* & \text{ if } Z = 0, \\ 2\theta - y^* & \text{ if } Z = 1. \end{cases}
\eea
Note that to simulate a $p$-value from a distribution with density \eqref{pval dens} one could equivalently first generate a random number
$y$ drawn from the distribution $N(\theta, \eta^2)$ under the above alternative and then compute the $p$-value as $p = 2 \Phi(- |y| / u)$.

Now, to generate a (one-sided) $p$-value for the null hypothesis of a constant weight function we proceed as follows:
\begin{enumerate}
\item As test statistic to assess constancy of a monotone weight function $w$ we choose $T = \min w$.
\item Compute estimates $\hat \theta_0$ and $\hat \sigma_0^2$ from the observed collection of $p$-values $p_1, \ldots, p_n$ in a standard random effects model.
Also compute the monotone weight function $\hat w_0$ based on this collection.
\item Draw samples $(p_{j1}, \ldots, p_{jn})$ of $p$-values for $j = 1, \ldots, M$ where $p_{ji}$ follows a distribution with density
$f(\cdot; \hat \theta_0, \hat \sigma_0, \sqrt{u_i^2 + \hat \sigma_0^2})$ for $i = 1, \ldots, n$. For each of these samples
also compute the monotone weight function $\hat w_j$. It is important to realize that these samples of $p$-values come, by construction, from the null model,
i.e assuming no selection bias.
\item Compute the test statistics $\hat T_0 = \min \hat w_0$ and $\hat T^{(j)} = \min \hat w_j$ for $j = 1, \ldots, M$.
\item The proposed approximate $p$-value that quantifies the evidence against a constant weight function is then
\bea
    p &=& \frac{1 + \#\{j \le M \ : \ \hat T_0 \le \hat T^{(j)}\}}{1 + M}.
\eea
\end{enumerate}
The function \code{DearBeggMonotonePvalSelection} implements this procedure in \pkg{selectMeta}.

\section{Examples} \label{examples}

\paragraph{Open classroom data}
As a first example, and to compare the monotone to the non-monotone approach of \cite{dear_92}, we re-analyze the famous
open classroom education data initially presented by \citet[p. 303]{hedges_85} and re-analyzed
by \cite{iyengar_88} and \cite{dear_92}. For convenience, the data is reproduced
in Table~\ref{tab: hedgesolkin} (compare \citealp[Table 4]{iyengar_88}).

All these studies assessed the effect of open vs. traditional education on student creativity, measured by some continuous quantity (in fact, we did not find
neither in \citealp{hedges_85} nor in \citealp{iyengar_88} the exact description of what was actually measured).
In Table~\ref{tab: hedgesolkin}, $N_i$ denotes the sample size in each of the two samples (so all the studies were perfectly balanced),
$y_i$ is the mean difference (the effect measure), $u_i$ are the standard errors, and $p_i$ are the computed $p$-values,
$p_i = 2 \Phi(-|y_i| / u_i)$.

\renewcommand{\baselinestretch}{1.2}

\renewcommand{\baselinestretch}{1.2}
\begin{table}[h]
\begin{center}
{\footnotesize
\begin{tabular}{rrrrr}
  \hline
$i$ & $N_i$ & $y_i$ & $u_i$ & $p_i$ \\
  \hline
1 & 10 &  0.081 & 0.45 & 0.86 \\
  2 & 10 &  0.308 & 0.45 & 0.49 \\
  3 & 39 & -0.178 & 0.23 & 0.43 \\
  4 & 50 & -0.234 & 0.20 & 0.24 \\
  5 & 10 &  0.598 & 0.45 & 0.18 \\
  6 & 22 &  0.563 & 0.30 & 0.06 \\
  7 & 40 &  0.535 & 0.22 & 0.02 \\
  8 & 36 &  0.779 & 0.24 & 0.0009 \\
  9 & 20 &  1.052 & 0.32 & 0.0009 \\
  10 & 90 & -0.583 & 0.15 & 0.0001 \\
   \hline
\end{tabular}
}
\caption{Studies of effects of open vs. traditional education on creativity.}
\label{tab: hedgesolkin}
\end{center}
\end{table}\renewcommand{\baselinestretch}{1}

\begin{figure}[h!]
\begin{center}
\vspace*{-0.8cm}
\setkeys{Gin}{width= 1\textwidth}
\includegraphics{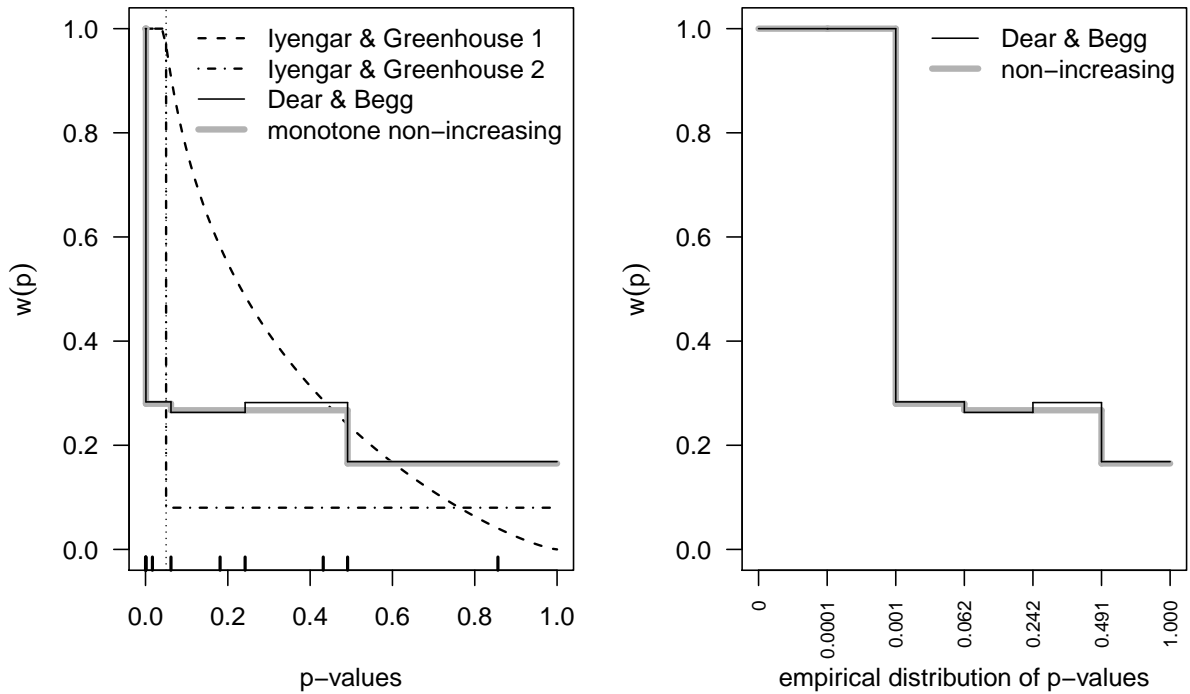}
\end{center}
\vspace*{-0.8cm}
\caption{Estimated weight functions for the open classroom education dataset.\label{fig: classroom}}
\end{figure}

In Figure~\ref{fig: classroom} we present the following estimates of the weight function:
\bi
\item The parametric weight functions $w_1$ and $w_2$ proposed in \citet[p. 113]{iyengar_88},
\item the nonparametric variant of \cite{dear_92},
\item and our new proposal: the nonparametric weight function constrained to be non-increasing.
\ei As in \cite{dear_92} we provide two plots: one with the original $p$-value scaling of the $x$-axis
and one where on the $x$-axis we plot the limits of the pairwise groups of $p$-values, where these limits are
equally spaced.
Note that in the latter plot (1) the parametric weight functions are not displayable and (2)
one must carefully study the horizontal axis to determine the $p$-values
represented by the estimated weight functions.
At the bottom of the first plot, we also indicated the observed 10 $p$-values with vertical ticks.

Since the nonparametric estimate of \cite{dear_92} is already ``almost'' non-increasing it comes without surprise that
the monotone estimate of $w$ is very similar to its unconstrained counterpart. The estimated weight functions clearly
indicate publication bias, an observation already made in \citet[p. 243]{dear_92}.
This is further supported by the $p$-value computed according to the procedure outlined in Section~\ref{quantifying evidence} that amounts to
$p_{\text{education}} = 0.096$ (for $M = 1000$ runs).
Now, having an estimate of $w$ at hand yields some more insight in the selection process:
According to the monotone estimate, the probability of a $p$-value that is larger than 0.001 to be published
is only 28.0\% compared to a $p$-value $\le 0.001$.


Finally, estimates for this dataset from our monotone selection model are $\hat \theta = 0.14$ and
$\hat \sigma^2 = 0.11$ with 95\% approximate profile likelihood confidence interval for $\theta$ of $[-0.08, 0.57]$. Compare these to estimates received from a standard random effects model that amount to $\hat \theta = 0.26$ and $\hat \sigma^2 = 0.30$
with estimated confidence interval for $\theta$ of $[-0.12, 0.65]$.
The adjustment for selection thus attenuates the effect estimate and narrows the confidence interval but does not change the conclusion about
significance of $\theta$ at a significance level of $\alpha = 0.05$.

\paragraph{Environmental tobacco smoke data}\label{tobacco}

\begin{figure}[h!]
\begin{center}
\vspace*{-0.8cm}
\setkeys{Gin}{width= 1\textwidth}
\includegraphics{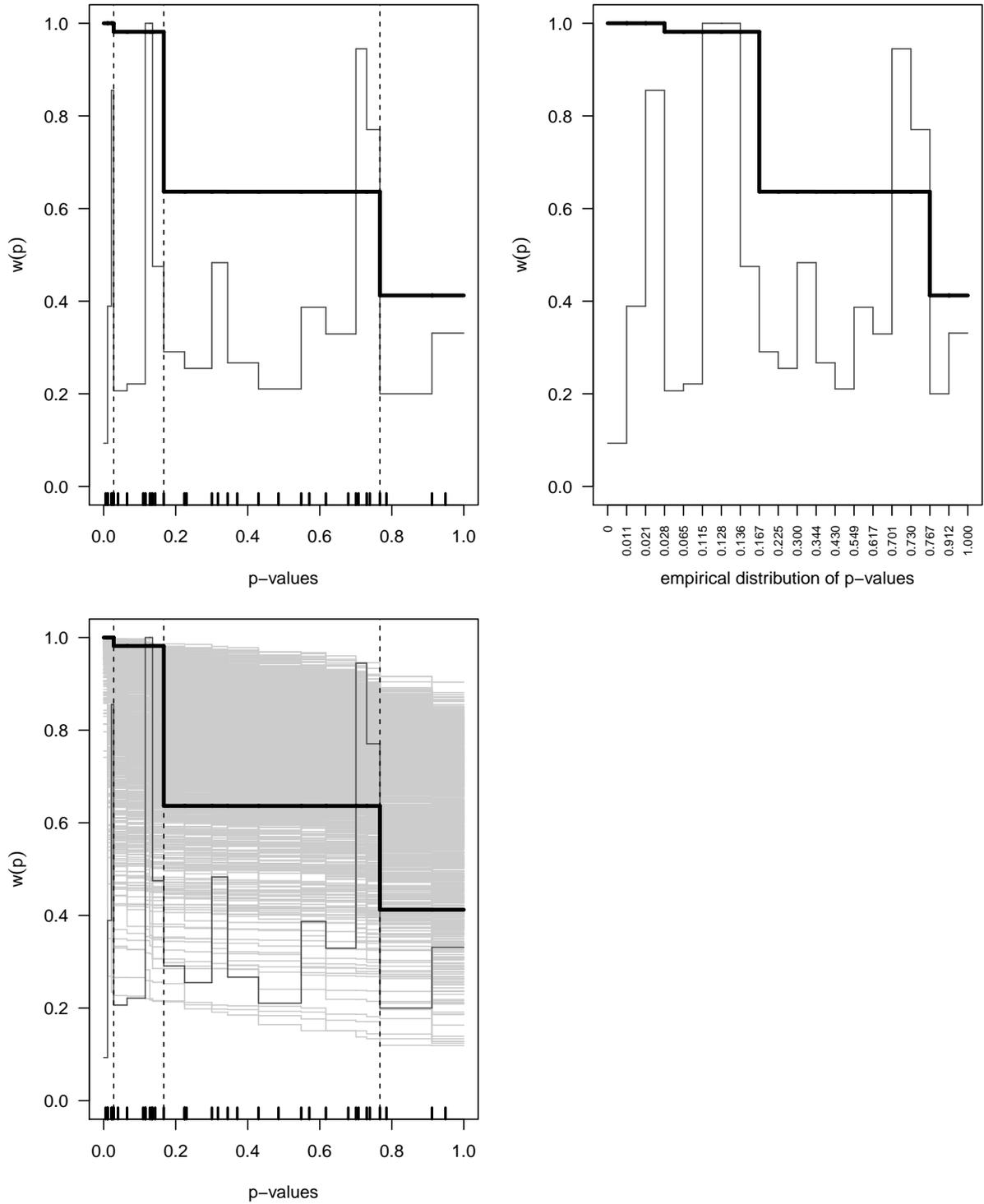}
\end{center}
\vspace*{-0.8cm}
\caption{Estimated weight functions for the environmental tobacco smoke dataset ($n = 37, k = 19$).
Thick lines: monotone non-increasing weight functions. Thin lines: Unrestricted weight functions.
Dashed vertical lines at 0.03, 0.17, 0.77.
\label{fig: tobacco}}
\end{figure}

In our second example we discuss a meta analysis \citep{hackshaw_97} of 37 studies concerned with the effect of
environmental tobacco smoke on lung-cancer in lifetime non-smokers. The effect in these studies is quantified via
the log relative risk.
Whether this meta analysis suffers from
publication bias has been a matter of ongoing controversy, see \citet[p. 91]{rothstein_05}. In the original publication
a peculiar form of ``failsafe $N$'' analysis was conducted and the authors concluded that there is no reason to suspect
publication bias. In a re-analysis however, \cite{copas_00} (see also the correspondence following that paper on
\url{www.bmj.com}), applying the method introduced in \cite{copas_99}, came to the conclusion that ``the possibility of
publication bias cannot be ruled out altogether, and at least some publication bias is needed to explain the trend
we found.'' However, neither the funnel plot, nor the method by \cite{copas_99}, or \cite{copas_08} yields real insight in the nature
of the selection process that may be at work. On the other hand, we can estimate the weight function via the unconstrained and the monotone approach,
see Figure~\ref{fig: tobacco}.

First, unlike claimed in \citet[comment to Figure 2]{dear_92}, note that from the
unconstrained estimate it is not evident whether publication bias is operating on this dataset.

Again, we can gain some insight in the possible selection mechanism by looking at the estimated weight function which
reveals an interesting pattern: One observes four distinct regions which are given by the
intervals $[0, 0.03]$, $(0.03, 0.17]$, $(0.17, 0.77]$, $(0.77, 1.00]$
where $w$ is constant. These regions are indicated with vertical dashed lines
in the left plot of Figure~\ref{fig: tobacco}. Not surprisingly, sharp drops in the weight function appear around
``psychological barriers'' for $p$-values, namely 0.05 and maybe 0.15. In passing we remark that discontinuities of the estimated weight function
can only happen at actually observed $p$-values.

The probability of selecting a study with $p$-value larger than 0.17 is only
64.8\% of that of one with a $p$-value at most
0.17.
In addition, the relatively small $p$-value of $p_\text{tobacco} = 0.13$ computed according the method
described in Section~\ref{quantifying evidence} reveals some evidence against a constant weight
function in Figure~\ref{fig: tobacco}. For these reasons it seems therefore plausible that publication bias is at work
here and we thus agree with the conclusion of \cite{copas_00} and \citet[p. 164]{hedges_05}.


Furthermore, in a standard random effects meta analysis model, we get estimates $\hat \theta = 0.21$ and
$\hat \sigma^2 = 0.02$ with estimated confidence interval for $\theta$ of $[0.12, 0.31]$.
These estimates are attenuated to $\hat \theta = 0.17$ and $\hat \sigma^2 = 0.01$
in the monotone selection model, with 95\% approximate profile likelihood confidence interval for $\theta$ of
$[0.08, 0.26]$.
These changes are very similar to those observed by \cite{hedges_05} when choosing their somewhat related weight function.
And the significant effect of environmental tobacco smoke on lung cancer persists after adjusting for selection bias.

Finally, to illustrate the computation of the suggested $p$-value, in the lower plot in Figure~\ref{fig: tobacco}, we have plotted in grey the estimated weight functions
for the $M = 1000$ samples generated under the assumption of no selection. This gives an impression what selection function can be expected
under no selection.


\section{Software and reproducibility} \label{software}
Although some of the methods discussed here have been around for some time, it seems as if none of them has found
its way in the daily routine of meta analysts. \citet[p. 439]{sutton_00} explain this lack of use by
``One of the explanations for this is almost certainly the complexity of many of
the approaches (particularly the selection models, and Copas' approach), and the lack
of user-friendly software available to implement them.'' In addition,
``A further reason for low penetration is possibly lack of acceptance of such methods.
and ``Selection models are quite sophisticated and there is currently a lack of software to implement them''.
To foster the use of selection models by meta analysts we have therefore implemented
\bi
\item the parametric weight functions $w_1$ and $w_2$ from \cite{iyengar_88} as well as maximum likelihood estimation of their corresponding parameters,
\item the nonparametric method of \cite{dear_92},
\item our new variant that provides a monotone version of the latter estimate, including estimation of the selection bias adjusted estimates
of $\theta$ and $\sigma$ as well as the approximate profile likelihood confidence interval for $\theta$,
\item the density, distribution, and quantile function as well as random number generation from the $p$-value density \eqref{pval dens},
\item and the procedure to compute a $p$-value to assess the null hypothesis of no selection introduced in Section~\ref{quantifying evidence},
\ei
in a new \proglang{R} package \pkg{selectMeta} \citep{selectMeta} which is available from \code{CRAN}.
In addition, we provide in \pkg{selectMeta} the two datasets analyzed in Section~\ref{examples}.

Making the software and datasets discussed in this paper accessible enables reproducibility
of the results and plots. The code that generates Figures~\ref{fig: classroom} and \ref{fig: tobacco} as well as the computation of
the $p$-value introduced in Section~\ref{quantifying evidence} for these two examples can be found in the help file
for the function \code{DearBegg} in \pkg{selectMeta}.

This document was created using \code{Sweave} \citep{leisch_02}, \LaTeX{} \citep{knuth_84, lamport_94}, \proglang{R} 2.12.2 \citep{R}
with the \proglang{R} packages
\pkg{selectMeta} (\citealp{selectMeta}, Version 1.0.3),
\pkg{DEoptim} (\citealp{DEoptim}, Version 2.0-9),
\pkg{meta} (\citealp{meta}, Version 1.6-1),
\pkg{reporttools} (\citealp{reporttools}, Version 1.0.5), and
\pkg{cacheSweave} (\citealp{peng_08}, Version 0.4-5).

\section{Final remarks} \label{discussion}
We propose and analyze a new type of monotone frequentist nonparametric weight functions as a visual tool
to gain insight in the study selection process when publication
bias in meta analysis must be suspected.
Selection models have not yet entered the standard toolbox of meta analysts, presumably due to lack of easy accessible
software. Our goal was to reduce this gap by collecting many existing and our new approach as functions in a new \proglang{R}
package \pkg{selectMeta} \citep{selectMeta}.

More research is necessary to popularize selection models. We intend to develop a smooth version of our new estimator
by imposing not only a monotonicity but also a smoothness constraint on the log-likelihood. However, already
difficult algorithmic aspects
are not facilitated by such additional regularization structure. We further plan to apply and adapt the method
of \cite{sun_97} to meta analysis.
Finally, \citet[Eq. 9.6]{hedges_05} describe how to incorporate $\theta$ not only as a simple
number, but rather depending on covariates in a regression model. This approach should also allow for a generalization
to other than the specific selection model they look at.

\paragraph{Acknowledgments}
I thank an external reviewer and the associated editor for valuable comments.

\paragraph{Conflict of Interest}
The author has declared no conflict of interest.

\begin{appendix}

\section{Derivation of relevant quantities} \label{wloglik}
In this brief appendix, we provide some additional computations that lead to the log-likelihood function,
merely for the reader's convenience. As a matter of fact and since the log-likelihood used in this paper is the one
of \cite{dear_92}, a derivation of $l$ can also be found there. However, here we try to be a bit more explicit.

Since the weight function $w$ is defined to be piecewise constant with values $w_j$, the normalizing constants $A_i$ simplify to
\bea
    A_i &=& \sum_{j=1}^k w_j \int_{y:w_i(y) = w_j} \phi\Bl(\frac{y-\theta}{\eta_i} \Br) \d y \\
    &=:& \sum_{j=1}^k w_j H_{ij}.
\eea The quantities $H_{ij}$ can be computed as follows:
\bea
    H_{ij} &=& \int_{y:w_i(y) = w_j} \phi\Bl(\frac{y-\theta}{\eta_i} \Br) \d y \\
    &=& \int_{b_{i, 2j-2} \le |y| < b_{i, 2j}} \phi\Bl(\frac{y-\theta}{\eta_i} \Br) \d y \\
    &=& \int_{b_{i, 2j-2}}^{b_{i, 2j}} \phi\Bl(\frac{y-\theta}{\eta_i} \Br) \d y +  \int_{-b_{i, 2j}}^{-b_{i, 2j-2}} \phi\Bl(\frac{y-\theta}{\eta_i} \Br) \d y \\
    &=& \Phi\Bl(\frac{b_{i, 2j}-\theta}{\eta_i}\Br)-\Phi\Bl(\frac{b_{i, 2j-2}-\theta}{\eta_i}\Br) + \Phi\Bl(\frac{-b_{i, 2j-2}-\theta}{\eta_i}\Br)-\Phi\Bl(\frac{-b_{i, 2j}-\theta}{\eta_i}\Br) \\
    &=& \Phi(a_{ij})-\Phi(b_{ij}) + \Phi(c_{ij})-\Phi(d_{ij})
\eea where we defined
\bea
    a_{ij} \ = \ \frac{u_i |y_{2j}|/u_{2j}-\theta}{\eta_i} &&  b_{ij} \ = \ \frac{u_i |y_{2j-2}|/u_{2j-2}-\theta}{\eta_i} \\
    c_{ij} \ = \ \frac{-u_i |y_{2j-2}|/u_{2j-2}-\theta}{\eta_i} && d_{ij} \ = \ \frac{-u_i |y_{2j}|/u_{2j}-\theta}{\eta_i},
\eea see \citet[Appendix]{dear_92}.
Consider the following ``boundary cases'': Defining $p_0 = 1$ and $p_{2k} = 0$, we get
$b_{i, 0} = 0$ and $b_{i, 2k} = \infty$, what immediately entails
\bea
    H_{i1} &=& \Phi\Bl(\frac{b_{i, 2j}-\theta}{\eta_i}\Br)-\Phi\Bl(\frac{-\theta}{\eta_i}\Br) + \Phi\Bl(\frac{-\theta}{\eta_i}\Br)-\Phi\Bl(\frac{-b_{i, 2j}-\theta}{\eta_i}\Br) \\
    &=&  \Phi\Bl(\frac{u_i |y_2|/u_2-\theta}{\eta_i}\Br)-\Phi\Bl(\frac{-u_i |y_2|/u_2-\theta}{\eta_i}\Br) \ = \ \Phi(a_{i1}) - \Phi(d_{i1}).
\eea On the other hand, 
\bea
    H_{ik} &=& \Phi\Bl(\frac{b_{i, 2k}-\theta}{\eta_i}\Br)-\Phi\Bl(\frac{b_{i, 2k-2}-\theta}{\eta_i}\Br) + \Phi\Bl(\frac{-b_{i, 2k-2}-\theta}{\eta_i}\Br)-\Phi\Bl(\frac{-b_{i, 2k}-\theta}{\eta_i}\Br) \\
    &=& 1-\Phi\Bl(\frac{u_i |y_{2k-2}|/u_{2k-2}-\theta}{\eta_i}\Br) + \Phi\Bl(\frac{-u_i |y_{2k-2}|/u_{2k-2}-\theta}{\eta_i}\Br) \ = \ 1-\Phi(b_{ik})+\Phi(c_{ik}).
\eea
Plugging all the above quantities into \eqref{eq: L}, the log-likelihood function amounts to
\bea
    l(\ve{w}, \theta, \sigma^2) &=& \log L(\ve{w}, \theta, \sigma^2) \nonumber \\
    &=& \sum_{i=1}^n \log w_i(y_i) + \sum_{i=1}^n \log \Bl\{\eta_i^{-1} \phi\Bl(\frac{y_i-\theta}{\eta_i}\Br)\Br\} -  \sum_{i=1}^n \log \Bl(\sum_{j=1}^k w_j H_{ij}\Br) \nonumber \\
    &=& -\frac{n}{2}\log(2\pi) + \sum_{j=1}^k \lambda_j \log w_j - \sum_{i=1}^n \log\eta_i  -\frac{1}{2}\sum_{i=1}^n \Bl( \frac{y_i-\theta}{\eta_i}\Br)^2 -  \sum_{i=1}^n \log A_i. \label{eq: loglik}
\eea

\end{appendix}



\bibliographystyle{ims}
\bibliography{metamonotone}

\end{document}